\documentclass[10pt]{article}
\usepackage{amsmath,amssymb}
\usepackage{graphicx}
 \allowdisplaybreaks
 \newcommand{\be}{\begin{equation}}
 \newcommand{\ee}{\end{equation}}
 \newcommand{\bea}{\begin{eqnarray}}
 \newcommand{\eea}{\end{eqnarray}}
 
 \newcommand{\nn}{\nonumber}
 
 \newcommand{\wtd}{\widetilde}
 \newcommand{\pd}{\partial}

 \newcommand{\cE}{{\cal E}}
 
 \newcommand{\cH}{{\cal H}}

 \newcommand{\cL}{{\cal L}}

 \newcommand{\cO}{{\cal O}}
 \newcommand{\cP}{{\cal P}}
 
 \newcommand{\cT}{{\cal T}}

 \def\bra#1{\left\langle#1\right|}
 \def\ket#1{\left|#1\right\rangle}
 \def\brk#1{\left\langle#1\right\rangle}

\long\def\symbolfootnote[#1]#2{\begingroup%
\def\thefootnote{\fnsymbol{footnote}}\footnote[#1]{#2}\endgroup}

\newcommand{\aei}{\it Max Planck Institute for Gravitational Physics
(Albert Einstein Institute)\\ Am M\"uhlenberg 1, 14476 Golm,
Germany}

\begin{document}
\thispagestyle{empty}
\begin{flushright}
\hfill{AEI-2013-205}
\end{flushright}
\begin{center}

~\vspace{20pt}

{\Large\bf Fluctuating Black Hole Horizons}

\vspace{25pt}

Jianwei Mei\symbolfootnote[1]{Email:~\sf jwmei@aei.mpg.de}

\vspace{10pt}\aei

\vspace{2cm}

\centerline{\bf Abstract}
\end{center}

In this paper we treat the black hole horizon as a physical
boundary to the spacetime and study its dynamics following from
the Gibbons-Hawking-York boundary term. Using the Kerr black hole
as an example we derive an effective action that describes, in the
large wave number limit, a massless Klein-Gordon field living on
the average location of the boundary. Complete solutions can be
found in the small rotation limit of the black hole. The
formulation suggests that the boundary can be treated in the same
way as any other matter contributions. In particular, the angular
momentum of the boundary matches exactly with that of the black
hole, suggesting an interesting possibility that all charges
(including the entropy) of the black hole are carried by the
boundary. Using this as input, we derive predictions on the Planck
scale properties of the boundary.

 \newpage

\tableofcontents

\section{Introduction}

By far our best understanding of black hole entropy has been based
on the idea of holography
\cite{'tHooft:1993gx,Susskind:1994vu,Bousso:2002ju}. In this
framework one assumes that quantum gravity in a given black hole
background has a dual description in terms of a field theory
defined on the boundary of the space
\cite{Brown:1986nw,Maldacena:1997re,Gubser:1998bc,Witten:1998qj}.
As far as black holes are concerned, all such dual descriptions
either have been deduced in string theory \cite{Strominger:1996sh}
or have arisen from considerations of asymptotic symmetries
\cite{Brown:1986nw,Strominger:1997eq}. For more realistic black
holes in General Relativity (GR), such dual descriptions are only
known in bits and pieces
\cite{Carlip:1998wz,Guica:2008mu,Castro:2010fd} and a full
understanding of the black hole entropy is still not available.

The idea of gauge/gravity duality is very profound and has had
wide applications (see, e.g.
\cite{Aharony:1999ti,Policastro:2001yc,Policastro:2002se,Ryu:2006bv,Hartnoll:2009sz}),
but alternative methods to the black hole entropy should also be
explored. In particular, even from the holography perspective one
expects that both sides of a dual pair should be equally good (not
considering the level of technical difficulty) in explaining the
same physics. For a black hole, this means that one also expects
to understand its entropy from the pure gravity side. Here a
central problem is to identify the correct physical entity
(entities) that is (are) responsible for the thermodynamical
properties of the black hole.

In this respect, 't Hooft \cite{'tHooft:1984re} has looked at
point particles as possible carriers of the black hole charges. He
noted that there should be a cut-off at a tiny distance away from
outside the horizon, so that the density of states of the
particles do not diverge. He then calculated the total energy and
the entropy of the particles. In order for the entropy to match
with that of the black hole, he noticed that the cutoff must be
\be\delta\sim\frac{\ell_p^2}{r_+}\,,\label{delta.tHooft}\ee
where $\ell_p=\sqrt{G_N}\approx 1.6 \times 10^{-35}m$ is the
Planck length and $r_+$ is the radius of the black hole horizon.
This cut off is referred to as the ``brick wall". 't Hooft's
calculation was proposed as a toy model and there are obvious
problems with it, including the fact that the final answer of the
entropy depends on the number of available particle species living
in the vicinity of the horizon. What's more, the physical origin
of the ``brick wall" is not entirely clear. An idea for addressing
this later problem is recently proposed in \cite{Arzano:2013tsa},
using an earlier idea from York \cite{York:1983zb}. York attempted
to understand the black hole entropy by including back-reactions
from the quantum fields to the black hole background, which leads
to an oscillating metric \cite{York:1983zb}.

In this paper, instead of looking at point particles, we want to
study if extended objects can play any role in the statistical
origin of black hole entropy. Extended objects like D-branes are
well known in string theory and have played a crucial role in the
first successful string calculation of black hole entropy
\cite{Strominger:1996sh}. Here we shall study in detail the
possible connection between the dynamics of extended objects and
the black hole entropy directly within the framework of GR.

For this purpose we need to look for brane-like objects which can
be possibly related to the black hole thermodynamics. In fact, a
candidate is readily available, which is nothing but the black
hole horizon. The idea of viewing the black hole horizon as a
membrane has been much explored in the membrane paradigm
\cite{Parikh:1997ma,Thorne:1986iy}.\footnote{The author thanks
Jos$\acute{e}$ Lemos for this point and for the related reference
\cite{Lemos:2011ic}. There has also been earlier effort in
deriving boundary effective actions (see, e.g.
\cite{Carlip:2005tz}), as we will do in this paper. The author is
indebted to the referee for this later point.} The most direct
reason for this possibility is that, if the horizon is a dynamical
object, then its dynamics must be described by a field theory
living on the world volume of the horizon. As a result, the
entropy of the system must be proportional to the area of the
horizon. Still, treating the black hole horizon as a physical
entity may appear odd at the first glance. Let's discuss some
further hint for this possibility.

Back in 1998 Carlip has shown that one can identify conformal
symmetries on the (stretched) horizon, which can then be used to
infer information about the entropy of the black hole
\cite{Carlip:1998wz}. This idea has been reinforced in recent
years by the proposal of the Kerr/CFT correspondence, which showed
how to identify the conformal symmetries by zooming in the near
horizon region of an extremal Kerr black hole \cite{Guica:2008mu}.
Later effort has further shown how conformal symmetries can be
identified on the horizons of generic stationary and axisymmetric
black holes in generic dimensions
\cite{Lu:2008jk}-\cite{Mei:2012wd}. This suggests, in particular,
that the quantum nature of a black hole might be captured by a
dual conformal field theory (CFT) living on the horizon, and this
should be a generic feature in all spacetime dimensions. If true,
this can be the best explanation to the problem of ``Universality"
\cite{Carlip:2007qh}.

Another suggestive hint comes from the classic $AdS_5/CFT_4$
correspondence \cite{Maldacena:1997re}, where the boundary Super
Yang-Mills theory is related to the dynamics of a stack of
D-branes. By analogy, in the pure gravity case the dual field
theory may also describe dynamics of extended objects like the
branes. Since from the previous paragraph we know that the dual
field theory of a black hole likely lives on the horizon, it is
tempting to ask if the horizon itself is part of the objects
described by the dual field theory. What's more, if this is true,
we suggest that this could offer a way to obtain, instead of a
full theory, an (or a partial contribution from the horizon to
the) effective action of the dual field theory. The idea for doing
this is based on the fact that the horizon acts as a boundary to
the black hole spacetime.

After it was understood that the singularity at a black hole
horizon can be removed by a coordinate transformation, it has been
widely accepted that there is no true singularity on the horizon
and one can pass through it without experiencing anything
dramatic. This latter point, however, is being challenged recently
\cite{Braunstein:2009my,Almheiri:2012rt}. Despite this, everyone
agrees that the horizon acts as a unidirectional membrane, i.e. a
causal boundary of the spacetime. This is at the basis of trying
to interpret the black hole entropy as an entanglement entropy
(see, e.g. \cite{Solodukhin:2011gn}).

With the conformal symmetries mentioned above, the horizon could
be more than just a causal boundary. For a two dimensional
statistical system, conformal symmetry usually arises at critical
points of second order phase transitions. Since the conformal
symmetries related to a black hole horizon are also those in two
dimensions, this hints at the intriguing possibility that the
horizon could also be a boundary separating two different phases
of (the material that is making up) the spacetime. Inspired by
this, we consider the possibility that the horizon is a {\it
physical} boundary to the spacetime.

If treated as a brane with zero depth, a {\it physical} boundary
contributes to the stress energy tensor with a delta function in
the direction normal to the boundary, which then leads to a step
function in the metric. This means that the metric inside the
horizon is significantly different from that of a usual black
hole. In the following, however, we will only need the part of the
metric that is outside the horizon.

When there is a boundary to the spacetime, the Einstein-Hilbert
action must be supplemented by the Gibbons-Hawking-York boundary
term \cite{York:1972sj,Gibbons:1976ue} so that the variational
principle can be well defined. In the presence of a physical
boundary, one can view the boundary term as the contribution from
the boundary to the total action. So it is reasonable to suggest
that the dynamics of the boundary is governed by the boundary
action.\footnote{If the bulk action is not Einstein-Hilbert, then
the boundary term should be modified accordingly. If a black hole
exists in this new theory, then the dynamics of the horizon should
be governed by the new boundary action.}

In the rest of the paper we study the dynamics of the horizon as
predicted by the Gibbons-Hawking-York boundary term. Since the
horizon is assumed to be a physical boundary to the spacetime, we
will use the words ``horizon" and ``boundary" interchangeably.

In section 2, we collect all formulae related to the boundary
action which we will need in later sections. In section 3, an
effective action is derived from the Gibbons-Hawking-York boundary
term in the background of a Kerr black hole. Instead of living
precisely on the dynamical horizon, this effective action will be
defined on the average location of the boundary. In section 4, all
classical solutions to the effective action are found in the small
rotation limit of the black hole. The system can then be quantized
and the complete spectrum is found. In section 5, we study
thermodynamical properties of the system. We show that the black
hole angular momentum is fully accounted for by that of the
boundary. This motivates us to assume that all charges of the
black hole are carried by the horizon. With this assumption one
can get predictions on the Planck scale properties of the
boundary. The paper ends with a short summary in section 6.

\section{Boundary action}

Suppose that the boundary is defined by the function $B(x)=0$,
then the Gibbons-Hawking-York boundary term is given by
\be S_B=\frac1{16\pi\ell_p^2}\int_B(d^{n-1}x)_\mu N^\mu\sqrt{-g}\;
K=\frac1{16\pi}\int_Bd^{n-1}x\sqrt{-h}\;K\,, \label{action.B}\ee
where $(d^{n-1}x)_{\mu_1\cdots \mu_p} =\frac1{(n-p)!p!}
\varepsilon_{\mu_1\cdots\mu_p\nu_1\cdots \nu_{n-p}}dx^{\nu_1}
\wedge\cdots\wedge dx^{\nu_{n-p}}$, $|\varepsilon|=1$, $n$ is the
dimension of the bulk spacetime, $N_\mu=\pd_\mu B/|\pd B|\,$,
$|\pd B|=\sqrt{g^{\varrho \sigma}\pd_\varrho B \pd_\sigma B}\,$,
$g$ is the bulk metric, $h$ is the induced metric on the boundary,
and $K$ is the extrinsic curvature
\be K=g^{\mu\nu}K_{\mu\nu}\,,\quad K_{\mu\nu}=\nabla_\mu N_\nu
+\nabla_\nu N_\mu\,.\ee
We will set $\ell_p=1$ for most part of the calculation, but will
restore it when needed. It is always possible to choose the bulk
coordinates $x^\mu\in\{r, x^i\,(i=1,\cdots,n-1)\}$ as such that
the boundary function is of the form $B=r-f(x^i)$. One can then
explicitly check that
\be \int_B(d^{n-1}x)_\mu=\int_Bd^{n-1}xN_\mu|\pd B|\,,\quad
\sqrt{-h}=|\pd B|\sqrt{-g}\;,\label{B.def}\ee
where the coordinates on $B=0$ are taken to be $x^i$, and we have
used the definition $h_{ij}dx^idx^j=g_{\mu\nu}dx^\mu dx^\nu|_{
B=0}$. The boundary action (\ref{action.B}) is determined up to a
constant term, but which will not be relevant for our following
calculations.

The total action is
\be S_{tot}=\frac1{16\pi}\int d^nx\sqrt{-g}\, (R-2\Lambda)
+S_B\,,\label{action.tot}\ee
where $R$ is the Ricci scalar in the bulk and $\Lambda$ is the
bulk cosmological constant, which can be zero. A variation of the
bulk metric leads to
\bea\delta S_{tot}&=&\frac1{16\pi}\int d^nx\sqrt{-g}\,\delta
g^{\mu\nu}\Big(R_{\mu\nu}-\frac{R-2\Lambda}2g_{\mu\nu}\Big)
+\delta S'_B\nn\\
\delta S'_B&=&\frac1{16\pi}\int_Bd^{n-1}x\sqrt{-h}\,\Big[\wtd
\nabla_\alpha (N_\beta\delta g^{\alpha\beta})+\wtd\nabla_\beta
(N_\alpha \delta g^{\alpha\beta})-\delta g^{\alpha\beta}\frac{
K}2g_{\alpha\beta} -\cP_{\alpha\beta}N^\mu\nabla_\mu\delta
g^{\alpha\beta}\Big]\nn\\
&&+\frac1{16\pi}\int_Bd^{n-1}x\sqrt{-h}\,N_\mu \Big(g_{\alpha
\beta}\nabla^\mu\delta g^{\alpha \beta} -\nabla_\nu\delta
g^{\mu\nu}\Big) \,,\label{vari.action.tot}\eea
where $\cP_{\alpha\beta}=g_{\alpha\beta}-N_\alpha N_\beta$ is the
projector onto the boundary and $\wtd\nabla_\alpha
=\cP_\alpha^\beta \nabla_\beta$. The first line of $\delta S'_B$
comes from $\delta S_B$, while the second line comes from varying
the bulk action. Here one observes a big difference between the
physics of a boundary and that of an isolated system, i.e. the
stress energy tensor of a boundary also receives contributions
from the bulk action. This is crucial to a consistent calculation
in the following.

One can combine the two lines of (\ref{vari.action.tot}) and find
\bea\delta S'_B&=&\frac1{16\pi}\int_Bd^{n-1}x\sqrt{-h}\,\Big[
\frac{\delta g^{\alpha\beta}}2\Big(\wtd\nabla_\alpha N_\beta
+\wtd\nabla_\beta N_\alpha-K\cP_{\alpha\beta}\Big)+\wtd
\nabla_\lambda(N_\beta\cP_\alpha^\lambda\delta g^{\alpha
\beta})\Big]\nn\\
&=&\frac1{16\pi}\int_Bd^{n-1}x\sqrt{-h}\;\frac{\delta
g^{\alpha\beta}}2\cT_{\alpha\beta}\,,\nn\\
\cT_{\alpha\beta}&=&\wtd\nabla_\alpha N_\beta +\wtd\nabla_\beta
N_\alpha-K\cP_{\alpha\beta}+2N_\beta \Big(\frac12 \cP^{\rho\sigma}
\wtd\pd_\alpha\cP_{\rho\sigma}-\wtd\pd_\alpha\ln\sqrt{-h}\Big)\,,
\label{vari.action.B}\eea
where we have used $\int_Bd^{n-1}x\wtd\pd_\lambda f^\lambda=0$
and\footnote{I thank Stefan Theisen for help in simplifying the
result.}
\be\int_Bd^{n-1}x\sqrt{-h}\,\wtd\nabla_\lambda f^\lambda=\int_B
dx^{n-1} x\sqrt{-h}\, f^\lambda\Big(\frac12 \cP^{\rho\sigma}
\pd_\lambda\cP_{\rho\sigma}-\wtd\pd_\lambda\ln\sqrt{-h}\Big)\,.
\label{bound.inte}\ee
The last term in (\ref{vari.action.B}) can be shown to vanish and
$\cT_{\alpha\beta}$ becomes the quasi-local stress tensor by Brown
and York \cite{Brown:1992br}. The stress energy tensor of the
boundary can be found as
\be T_{\mu\nu}=-\frac2{\sqrt{-g}}\frac{\delta S'_B}{\delta
g^{\mu\nu}}=-\frac{\delta(B)|\pd B|}{16\pi}\cT_{\mu\nu}\,,
\label{Tuv.def}\ee
where we have used
\be\int_Bd^{n-1}x\sqrt{-h}\;=\int d^nx\delta(B)\sqrt{-h}\; =\int
d^nx\sqrt{-g}\;\delta(B) |\pd B|\,.\ee
The minus sign in the definition of $T_{\mu\nu}$ means that the
boundary is treated as a matter contribution. As mentioned before,
there is also a delta function in $T_{\mu\nu}$.

In a stationary and axisymmetric background, given the canonical
time $t$ and azimuthal angle $\phi$, the energy and the angular
momentum of the boundary are
\bea H=-\frac1{16\pi}\int_Bd^{n-1}x\sqrt{-h}\;\cT^t_t\,,\quad
J_\phi=-\frac1{16\pi}\int_Bd^{n-1}x\sqrt{-h}\;\cT^t_\phi\,.
\label{def.HJ}\eea

\section{Effective action in the Kerr background}

Let's now study (\ref{action.B}) in the background of a Kerr black
hole. The metric is
\bea ds^2&=&f\Big(\frac{dr^2}\Delta-\frac\Delta{v^2}dt^2\Big)
+\frac{fdx^2}{1-x^2}+\frac{v^2(1-x^2)}f(d\phi-w dt)^2\,,\nn\\
\Delta&=&(r-r_+)(r-r_-)\,,\quad w=\frac{r}{v^2}
\sqrt{r_+r_-}\,(r_++r_-) \,,\nn\\
f&=&r^2+r_+r_-x^2\,,\quad v^2=(r^2+r_+r_-)^2-\Delta\,
r_+r_-(1-x^2)\,,\label{metric.kerr4d}\eea
where $r_\pm=M\pm \sqrt{M^2-J^2/M^2}\,$ with $M$ and $J$ being the
mass and angular momentum of the black hole, respectively. The
outer (inner) horizon of the black hole is given by $r_+\,(r_-)$.
The black hole temperature, angular velocity and entropy are
\be T=\frac{r_+-r_-}{4\pi r_+(r_++r_-)}\,,\quad \Omega=\frac{
\sqrt{r_-r_+}}{r_+(r_++r_-)}\,,\quad S=\pi r_+(r_++r_-)\,.\ee
As the radius $r\to\infty$, the metric (\ref{metric.kerr4d})
approaches that of a Minkowski spacetime, where $t$ is the time,
$x\in[-1,1]$ and $\phi=\phi+2\pi$. We will refer to these as the
canonical coordinates. The determinant of the metric
(\ref{metric.kerr4d}) is $\sqrt{-g}=f$. In the background of
(\ref{metric.kerr4d}), the extrinsic curvature for a surface with
normal vector $N_\mu$ is
\be K=\frac2f\Big\{\pd_r(\Delta N_r)+\pd_x[(1-x^2)N_x]+\frac{f^2
\pd_\phi N_\phi}{v^2(1-x^2)}-\frac{v^2}\Delta(\pd_t +w\pd_\phi)
(N_t+wN_\phi)\Big\}\,. \ee

As will be justified later, the horizon described by
(\ref{action.B}) fluctuates around an average location in the
background of (\ref{metric.kerr4d}). The fluctuating horizon
carries part of the black hole energy,\footnote{We will suggest
later that the boundary actually carries all the energy of the
black hole.} and this energy is distributed to each excitation of
the boundary. So each excitation can be viewed as living in a
background that has an energy slightly smaller than that of the
full black hole. In canonical coordinates, this effectively means
that the physically fluctuating horizon always lives {\it outside}
the coordinate singularity of the background metric. For later
convenience, let's call the horizon of the background metric, such
as the $r_+$ in (\ref{metric.kerr4d}), the ``background horizon",
while the physically fluctuating horizon simply the ``horizon" or
the ``boundary", interchangeably.

With the fixed background (\ref{metric.kerr4d}), one can then
assume that the boundary is centered at $r_0=r_+(1+ \epsilon)$ and
is fluctuating with an amplitude $r_+\epsilon\,|\Phi(x,\phi,t)|$,
where $\epsilon>0$ is a small parameter. The configuration
function of the fluctuating horizon is then given by
\be B=r-r_+\Big\{1+\epsilon\Big[1+\Phi(x,\phi,t)
\Big]\Big\}=0\,.\label{def.B}\ee
The unit normal vector is
\bea&&N_r=\frac1{|\pd B|}\,,\quad N_i=-\epsilon\, r_+
\frac{\pd_i\Phi}{|\pd B|}\,,\quad i=x,\phi,t\,,\nn\\
&&|\pd B|^2=\frac\Delta{f}+(\epsilon\,r_+)^2\Big\{\frac{1-x^2}f
(\pd_x\Phi)^2 +\frac{f(\pd_\phi\Phi)^2}{v^2 (1-x^2)}\nn\\
&&\qquad\qquad\qquad\qquad\quad-\frac{v^2}{f\Delta}[(\pd_t
+w\pd_\phi)\Phi]^2\Big\}\,.\eea
The induced metric on the boundary is
\bea ds_H^2&=&-\frac{f\Delta}{v^2}dt^2 +\frac{fdx^2}{1-x^2}
+\frac{v^2(1-x^2)}fd\phi^2\nn\\
&&+(\epsilon r_+)^2\frac{f}\Delta\Big(\pd_x\Phi dx +\pd_\phi\Phi
d\phi+\pd_t\Phi dt\Big)^2\,, \label{metric.H}\eea
which has the determinant $\sqrt{-h}=f\,|\pd B|=|\pd B|
\sqrt{-g}\,$, as is expected.

By definition, one plugs (\ref{def.B}) - (\ref{metric.H}) into
(\ref{action.B}) and then let $r=r_+\{1+\epsilon[1+ \Phi(x,\phi,
t)]\}$ to obtain the action on the boundary. As we will see later,
$\epsilon$ is an extremely small parameter. So one can firstly
expand the action (\ref{action.B}) around $\epsilon$ and then look
at the weak field limit $|\Phi|\to0$. After doing this for
(\ref{action.B}), however, we find that
\bea S_B&=&\frac{r_+-r_-}4-\frac\mu2\int_Bdxd\phi dt\,\Big[\cL_0
+\frac{2r_+ (r_++r_-)^2\Phi(\pd_t+\Omega\pd_\phi)^2\Phi}{r_+-r_-}
+\cO( \Phi^3)\Big]+\cO(\sqrt\epsilon\,)\,,\nn\\
\cL_0&=&\frac{2r_+(r_++r_-)^2}{r_+-r_-}[(\pd_t+\Omega\,
\pd_\phi)\Phi]^2-(1-x^2)(\sqrt\epsilon\;\pd_x\Phi)^2
-\frac{(r_++r_- x^2)^2(\sqrt\epsilon\;\pd_\phi\Phi)^2}{
(r_++r_-)^2(1-x^2)}\,, \label{def.L0}\eea
where $\mu=r_+/(8\pi)$ and we have thrown away all boundary terms.
The minus sign in front of the $\cL_0$ integral is consistent with
that the boundary should be treated as a matter contribution, as
was assumed in (\ref{Tuv.def}). The derivatives $\pd_x$ and
$\pd_\phi$ only appear in $\cL_0$ through the combinations
$\sqrt\epsilon\,\pd_x\Phi$ and $\sqrt\epsilon\,
\pd_\phi\Phi$.\footnote{Note one can always let $\pd_t+\Omega
\pd_\phi\to\pd_t$ by a coordinate redefinition $\phi\to\phi+\Omega
t$.} Given the smallness of $\epsilon$, this means that only the
large wave number modes can have significant contributions.

The $(\pd_t+\Omega\pd_\phi)$-terms in $S_B$ consist a total
derivative and drop out of the integral, which means that the
action $S_B$ does not predict any interesting dynamics of the
boundary. Normally this would be the end of the story. But then we
notice an interesting feature. That is, instead of
$r=r_+\{1+\epsilon[ 1+\Phi(x,\phi,t)]\}$, if we look at the
average location of the brane, $r=r_0=r_+(1+\epsilon)$, we indeed
find well defined dynamics,
\be S_H=-\frac{r_+-r_-}4+\frac\mu2\int_Hdxd\phi dt\Big[\cL_0
+\cO(\Phi^3)\Big]+\cO(\sqrt\epsilon\,)\,,\label{action.B.r0}\ee
where the subscript $H$ means the integral is taken over the
surface $r_0=r_+(1+\epsilon)$. We have included an extra minus
sign in the boundary action so that $S_H$ is in the proper form of
a matter contribution.

One may ask if there is any physical reason to believe in
(\ref{action.B.r0}). A possible explanation comes from the
approximation that we have made when deriving (\ref{def.L0}) and
(\ref{action.B.r0}). When the boundary fluctuates, the metric near
the boundary will fluctuate accordingly. But for technical reasons
we have to rely on the average field approximation and use the
average metric (\ref{metric.kerr4d}), instead of the real time
metric, in the calculations. Although a concrete proof is hard to
get at the moment, this approximation suggests that it could make
more sense that we also derive the effective action at the average
location of the boundary, just as in (\ref{action.B.r0}). In any
case, this is still a weak link in the whole construction that we
wish to improve in future works.

For the moment, we make the extra assumption that
(\ref{action.B.r0}) is the correct effective action for the
boundary. (Hopefully we can eliminate the need for this extra
assumption in the near future!) And we will show in the following
that this action does lead to physically reasonable results.

For the charges (\ref{def.HJ}) the situation is similar. There is
no dynamics at $r=r_+[1+\epsilon(1+\Phi)]$, while at $r=r_0$ we
find
\bea H&=&-\Omega J_\phi+\frac\mu2\int_Hdxd\phi dt\Big\{\frac{2r_+
(r_++r_-)^2}{r_+-r_-}[(\pd_t+\Omega\,\pd_\phi)
\Phi]^2-\cL_0\Big\}\,,\label{H.rst1}\\
J_\phi&=&\frac\mu2\int_Hdxd\phi dt\Big\{\frac{2r_+(r_++r_-)^2
}{r_+ -r_-}(\pd_t+\Omega\,\pd_\phi)\Phi\,\pd_\phi\Phi\nn\\
&&\qquad+\Omega\frac{(r_++r_-)^2(1-x^2)}{(r_++r_-x^2)^2}
\Big[3r_+^2 -r_-^2x^2+r_+r_-(1+x^2)\Big] \Big\}\nn\\
&=&J+\frac\mu2\int_Hdxd\phi dt\Big[\frac{2r_+(r_++r_-)^2}{r_+
-r_-}(\pd_t+\Omega\,\pd_\phi) \Phi\,\pd_\phi\Phi\Big]\,,
\label{J.rst1}\eea
where $J=\frac12(r_++r_-)\sqrt{r_+r_-}\,$ is the angular momentum
of the black hole.

Not considering the constant terms, one can infer from
(\ref{J.rst1}) that the canonical momentum conjugating to $\Phi$
is
\be\Pi_\Phi=\mu\frac{r_+(r_++r_-)^2}{r_+ -r_-}(\pd_t+\Omega\,
\pd_\phi)\Phi\,. \label{def.PiPhi}\ee
From (\ref{H.rst1}) one can also read off the Hamiltonian density
\bea\cH&=&-\Omega\Pi_\Phi\pd_\phi\Phi+\frac\mu2\Big\{\frac{2r_+
(r_++r_-)^2}{r_+-r_-}[(\pd_t+\Omega\,\pd_\phi)\Phi]^2-\cL_0\Big\}\nn\\
&=&\frac\mu2\Big\{\frac{2r_+ (r_++r_-)^2}{r_+-r_-} (\pd_t+\Omega\,
\pd_\phi) \Phi\pd_t\Phi -\cL_0\Big\}\,,\eea
which is in the familiar form $\cH=\Pi_\Phi\pd_t\Phi
-\frac12\mu\cL_0$.

For an isolated system, we expect $\Pi_\Phi=\frac{\delta
S_H}{\delta(\pd_t\Phi)}$. For a boundary, however, the
contribution from the bulk action modifies the relation. In the
present case, we actually find
\be\Pi_\Phi=\frac12\frac{\delta S_H}{\delta(\pd_t\Phi)}\,.\ee
This unusual relation will not cause any problem for our
calculations in the following.

\section{Classical solutions and quantization}

Let's now look at the solutions of (\ref{action.B.r0}). Instead of
considering the general case, lets focus on the small rotation
limit ($\Omega\to0\,\Rightarrow\,\rho\equiv r_-/r_+\to0$). In this
case,
\bea \cL_0&=&2r_+^2(1+3\rho)[(\pd_t+\Omega\,\pd_\phi)
\Phi]^2-(1-x^2)(\sqrt\epsilon\;\pd_x\Phi)^2\nn\\
&&-\Big(\frac1{1-x^2} -2\rho\Big)(\sqrt\epsilon\;
\pd_\phi\Phi)^2\,,\eea
where we have preserved terms up to the subleading order in
$\rho\to0$. The Hamiltonian and the angular momentum are
\bea H&=&-\Omega J_\phi+\frac\mu2\int_Hdxd\phi\Big\{(1-x^2)
(\sqrt\epsilon\;\pd_x\Phi)^2+\Big(\frac1{1-x^2} -2\rho\Big)
(\sqrt\epsilon\;\pd_\phi \Phi)^2\Big\}\,,\label{H.rst2}\\
J_\phi&=&J+\int_Hdxd\phi\,\mu\,r_+^2(1+3\rho)(\pd_t
+\Omega\,\pd_\phi)\Phi\,\pd_\phi\Phi\,.\label{J.rst2}\eea

With $\Phi=f_\ell^m(x)\exp\{i[m(\phi-\Omega t)-\cE_{\ell,m}t]\}$,
the equation of motion  from (\ref{action.B.r0}) is
\be2r_+^2(1+3\rho)\cE_{\ell,m}^2 f_\ell^m +\epsilon\,\pd_x\Big[
(1-x^2)\pd_xf_\ell^m\Big]-\Big(\frac1{1-x^2}-2\rho\Big)\epsilon\,
m^2f_\ell^m =0\,.\ee
This equation can be solved by the associated Legendre polynomials
$f_\ell^m=P_\ell^m(x)$, with
\bea\cE_{\ell,m}&=&\sqrt{\epsilon\,\frac{\ell(\ell+1)-2m^2\rho}{2
r_+^2(1+3\rho)}}\approx\frac\ell{r_+}\sqrt{\frac\epsilon2}\;
\Big[1-\rho\Big(\frac32+\frac{m^2}{\ell^2}\Big)\Big]\,,\nn\\
\ell&=&0,1,\cdots,\infty\,,\quad m=-\ell,\cdots,\ell\,.
\label{spectrum}\eea
(Here and in the following, when making approximations, we always
preserve terms up to the subleading order in $\rho\to0$ and up to
the leading order in $|m|\sim\ell\to\infty$.) The full solution
can be expanded as $\Phi=\sum_{\ell,m} \Phi_\ell^m$, where
\bea\Phi_\ell^m&=&N_\ell^mP_\ell^m(x)\Big\{a_{\ell,m}e^{i[m
(\phi-\Omega t)-\cE_{\ell,m}t]}+a_{\ell,m}^\dagger e^{-i[m
(\phi-\Omega t)-\cE_{\ell,m}t]}\Big\}\,,
\label{classical.Phi}\\
N_\ell^m&=&\sqrt{\frac1{2\mu\,r_+^2(1+3\rho)\cE_{\ell,m}}
\frac{2\ell+1}{4\pi}\frac{(\ell-m)!}{(\ell+m)!}}\;.
\label{def.Nlm}\eea

In quantization, one promotes $a_{\ell,m}$ and
$a_{\ell,m}^\dagger$ to operators $\hat{a}_{\ell,m}$ and
$\hat{a}_{\ell,m}^\dagger$, yielding $\hat\Phi$ and
$\hat\Pi_\Phi$. Then one imposes the following equal time
commutation relations,
\bea[\hat\Phi(x,\phi,t),\hat\Pi_\Phi(x',\phi',t)]&=&i\delta(x-x')
\delta(\phi-\phi')\,,\nn\\
~[\hat\Pi_\Phi(x, \phi,t),\hat\Pi_\Phi(x',\phi',t)]&=&[\hat\Phi
(x,\phi,t),\hat\Phi(x',\phi',t)]=0\,, \label{com.basic}\eea
which, with (\ref{classical.Phi}) and (\ref{def.Nlm}), lead to
\be[\hat{a}_{\ell,m}\,,\,\hat{a}_{p,q}^\dagger]=\delta_{\ell,
p}\,\delta_{m,q}\,,\quad[\hat{a}_{\ell,m}\,,\,\hat{a}_{p,q}]
=[\hat{a}_{\ell,m}^\dagger\,,\,\hat{a}_{p,q}^\dagger]=0\,.
\label{com.basic1}\ee
As usual, $\hat{a}_{\ell,m}$ annihilates the vacuum state in the
Fock space, $\hat{a}_{\ell,m}\ket0=0$, and $\hat{a}_{\ell,
m}^\dagger$ generates multiparticle states, $\ket{N_{\ell,m}}\sim
(\hat{a}_{\ell, m}^\dagger)^{N_{\ell,m}}\ket0$, which have the
following properties
\be\hat{N}_{\ell,m}\ket{N_{\ell',m'}} =\delta_{\ell\ell'}
\delta_{mm'}N_{\ell,m}\ket{N_{\ell,m}} \,,\quad
\brk{N_{\ell,m}|N_{\ell', m'}}=\delta_{\ell\ell'}
\delta_{mm'}\,.\label{state.Nlm}\ee
Here $\hat{N}_{\ell,m}=\hat{a}_{\ell,m}^\dagger\hat{a}_{\ell, m}$,
and $N_{\ell,m}$ is a non-negative integer standing for the number
of corresponding particles. The Hamiltonian (\ref{H.rst2}) and the
angular momentum (\ref{J.rst2}) become
\bea\hat{H}&=&-\Omega\hat{J}_\phi+\sum_{\ell,m}\cE_{\ell,m}
\Big(\hat{N}_{\ell,m}+\frac12\hat{N}_0\Big)\nn\\
&=&\Big(-\Omega J+\sum_{\ell,m}\frac{\cE_{\ell,m}}2\Big)
\hat{N}_0+\sum_{\ell,m}\cE'_{\ell,m}\hat{N}_{\ell,m}\,,
\label{H.rst3}\\
\hat{J}_\phi&=&J\hat{N}_0-\sum_{\ell,m}\,m\hat{N}_{\ell,m}\,,
\label{J.rst3}\eea
where $\cE'_{\ell,m}=\cE_{\ell,m}+m\Omega$, and $\hat{N}_0$ is the
number operator for the vacuum state,
\be\hat{N}_0\ket0=\ket0\,,\quad \hat{N}_0\ket{N_{\ell,m}}=0\,.\ee
We have made the presence of $\hat{N}_0$ explicit so that both
(\ref{J.rst3}) and (\ref{H.rst3}) are well defined operator
equations. The fact that $\hat{H}$ and $\hat{J}_\phi$ have the
expected structures supports the result for the stress energy
tensor (\ref{Tuv.def}) and the result for the canonical momentum
(\ref{def.PiPhi}).

When there is gravity, we expect every piece of the Hamiltonian to
contribute to the total energy. That is why we keep both the
constant term and the terms of the zero point energy in
(\ref{H.rst3}). In order for the contribution from the zero point
energy to be finite, there must be a cutoff on the physically
available modes. As mentioned before, the existence of two
dimensional conformal symmetries on the horizon hints at the
possibility that the horizon is a boundary separating two
different phases of the spacetime. This can be taken as indicating
the existence of substructures of the spacetime. In this case, the
presence of a cutoff (say $N_c$) is very natural, which is
directly related to the minimal lattice spacing (say $a$) of the
substructures along the direction of the boundary,
\be N_c=\ell_{max}=m_{max}\approx\frac{2\pi r_+}{a}\,.
\label{def.Nc}\ee
With the cutoff, the contribution from the zero point energy is
\bea M_0&=&\sum_{\ell=0}^{N_c}\sum_{m=-\ell}^\ell
\frac{\cE_{\ell,m}}2
\approx\frac{N_c^3}{3r_+}\sqrt{\frac\epsilon2}\;
\Big(1-\frac{11\rho}6\Big)\,.\label{M0.def}\eea

\section{Statistics}

The thermal state of our scalar system with temperature $T$ and
angular velocity $\Omega$ is
\bea\ket\Psi&=&\ket0+\sum_{\ell,m}\sum_{N_{\ell,m}}\Big(\frac{
e^{-N_{\ell,m} (\beta \cE'_{\ell,m} +\alpha m)}}{\Xi_{\ell,m}}
\Big)^{1/2}\ket{N_{\ell,m}}\nn\\
&=&\ket0+\sum_{\ell,m}\sum_{N_{\ell,m}}\Big(\frac{e^{-N_{\ell,m}
\beta\cE_{\ell,m}}}{\Xi_{\ell,m}}\Big)^{1/2}\ket{N_{\ell,m}}\,,
\label{Psi.def} \eea
where $\Xi_{\ell,m}=\sum_{N_{\ell,m}}e^{-N_{\ell,m}\beta\cE_{
\ell,m}}=1/(1-e^{-\beta\cE_{\ell,m}})$, $\beta=1/T$ and $\alpha
=-\beta\Omega$. (The minus sign in $\alpha$ is due to the fact
that $\beta\Omega$ is a chemical potential.) By definition, the
thermal sate (\ref{Psi.def}) is space and time independent, and it
assigns the expected occupancy number to each particle (including
the vacuum),
\be\brk{\Psi|\hat{N}_0|\Psi}=1\,,\quad \brk{\Psi|\hat{N}_{\ell,
m}|\Psi}=\sum_{N_{\ell,m}}\frac{e^{-N_{\ell,m} \beta\cE_{\ell,
m}}}{ \Xi_{\ell,m}} N_{\ell,m}\,.\label{Psi.prop}\ee

As shown in (\ref{spectrum}), $\cE_{\ell,m}$ is an even function
in $m$. So ``particles" with $m>0$ and $m<0$ are evenly excited.
One can then derive from (\ref{H.rst3}), (\ref{J.rst3}) and
(\ref{Psi.prop}) that
\bea E&=&\bra\Psi\hat{H}\ket\Psi=-\Omega J+\sum_{\ell,m}
\frac{\cE_{\ell, m}}2 +\sum_{\ell,m}\sum_{N_{\ell,m}}
\cE'_{\ell,m} N_{\ell,m}\frac{e^{-N_{\ell,m}
\beta\cE_{\ell,m}}}{\Xi_{\ell,m}}\nn\\
&=&-\Omega J+\sum_{\ell,m}\frac{\cE_{\ell, m}}2
-\pd_\beta\sum_{\ell,m}\ln\Xi_{\ell,m}\,,\nn\\
J_\phi&=&\brk{\Psi|\hat{J}_\phi|\Psi}=J\,,
\label{prediction.J}\eea
where the second line for $E$ is possible because the terms linear
in $m$ cancel among themselves in $\cE'_{\ell,m}$. The result for
$J_\phi$ suggests that all the angular momentum of the black hole
is carried by the boundary, i.e., the horizon.

Now the key quantity to calculate is the the partition function,
\be\ln\Xi=\sum_{\ell,m}\ln\Xi_{\ell,m}=-\sum_{\ell=0}^{N_c}
\sum_{m=-\ell}^\ell\ln\Big (1-e^{-\beta\cE_{\ell,m}}\Big)\,.
\label{def.lnX}\ee
An explicit result is possible in the small rotation limit. In
this limit the black hole thermodynamical quantities can be
expanded as
\be M\approx\frac{r_+}2(1+\rho)\,,\quad J\approx \frac{r_+^2}2
\sqrt\rho\, (1+\rho)\,,\quad \Omega\approx\frac{\sqrt\rho}{r_+}
(1-\rho)\,,\quad T\approx\frac{1-2\rho}{4\pi r_+}\,,\quad
S\approx\pi r_+^2(1+\rho)\,.\ee
We firstly expand (\ref{def.lnX}) around $\rho\to0$, assuming that
$\beta$ is an unknown constant. Then we sum over $m$ and replace
$\sum_{\ell=0}^{N_c}$ by an integral. The result is (note $y=kn$
and $k=\frac\beta{r_+}\sqrt{\frac\epsilon2}\,$)
\bea\ln\Xi&\approx&-\frac2{k^2}\int_0^{kN_c}dyf_1(y)=-\frac2{k^2}
\Big[f_2(kN_c)-f_2(0)\Big]\,,\nn\\
f_1(y)&=&f'_2(y)=y\ln(1-e^{-y})-\frac{11\rho y^2}{6(e^y-1)}\,,
\quad f_1(0)=0\,,\nn\\
f_2(y)&=&yLi_2(e^{-y})+Li_3(e^{-y})+\frac{11\rho}3f_3(y)\,,
\quad f_2(0)=\zeta(3)\Big(1+\frac{11\rho}3\Big)\,,\nn\\
f_3(y)&=&yLi_2(e^{-y})+Li_3(e^{-y})-\frac{y^2}2\ln(1-e^{-y})\,,
\quad f_3(0)=\zeta(3)\,,\eea
where $Li_s(z)$ is the polylogarithm. The total energy and the
entropy are
\bea E&=&-\Omega J+M_0-\pd_\beta\ln\Xi\nn\\
&\approx&-\frac{r_-}2+\frac{N_c^3}{3r_+}\sqrt{\frac\epsilon2}\;
\Big(1-\frac{11\rho}6\Big)+\frac{k}\beta\Big\{\frac2{k^2}f_1
(kN_c)N_c-\frac4{b^3}\Big[f_2(kN_c)-f_2(0)\Big]\Big\}\nn\\
&=&\frac{\ell_p^2}{4\pi^3r_+^2\epsilon_0}\Big[\zeta(3)
-f_3(c_0)+\frac{c_0^3}{12}+\cO(\rho)\Big]M\,,\label{ratio.M}\\
S'&=&(1-\beta\pd_\beta)\ln\Xi\nn\\
&\approx&-\frac2{k^2} \Big[f_2(kN_c)-f_2(0)\Big]+k\Big\{\frac2{
k^2}f_1(kN_c)N_c-\frac4{k^3}\Big[f_2(kN_c)-f_2(0)\Big]\Big\}\nn\\
&=&\frac{3\ell_p^2}{4\pi^3r_+^2\epsilon_0}\Big[\zeta(3)
-f_3(c_0)-\frac{c_0^2}6\ln(1-e^{-c_0})+\cO(\rho)\Big]S\,,
\label{ratio.S}\eea
where we have let $\epsilon =\epsilon_0+\epsilon_1\rho
+\cO(\rho^2)$, $N_c=N_0+N_1\rho+\cO(\rho^2)$, and $c_0=2\pi\sqrt{2
\epsilon_0}\,N_0$. We have also written explicitly the Planck
length $\ell_p$ so that the ratios $\frac{E}M$ and $\frac{S'}S$
are obviously dimensionless.

In our setup, the parameter $\epsilon$ and the cutoff $N_c$ are
{\it a priori} not known. From (\ref{prediction.J}) we see that
the boundary carries all the angular momentum of the black hole.
This hints at an interesting possibility that all other charges of
the black hole are also carried by the horizon. We use this to get
an idea on what the values of $N_c$ and $\epsilon$ might be.

From the mass and the entropy there are two equations, $E=M$ and
$S'=S$. With two free parameters a solution is then
possible.\footnote{This point is not so trivial as it looks. For
example, if we do not include the contribution from the zero point
energy, it is then NOT possible to satisfy both $E=M$ and $S'=S$
simultaneously.} This can be done order by order in the expansion
of the small rotation limit. Up to the subleading order, we find
\be\sqrt{\epsilon_0}\,\approx\frac{\ell_p}{7.27r_+}\,,\quad
N_0\approx\frac{2\pi r_+}{2.84\ell_p}\,,\quad
\frac{\epsilon_1}{\epsilon_0}\approx0.22\,,\quad
\frac{N_1}{N_0}\approx1.61\,.\ee
A few comments are in order:
\begin{itemize}
\item Firstly, the putative lattice spacing introduced in
(\ref{def.Nc}) is
\be a\approx2.84\ell_p\,,\label{a.rst}\ee
which is independent of the properties of the black hole.

\item Secondly, $r_0$ differs from the background horizon $r_+$ by
\be\delta\approx r_+\epsilon_0\sim\frac{\ell_p^2}{r_+}\,,\ee
which is qualitative the same as (\ref{delta.tHooft}).  From
(\ref{metric.kerr4d}), the physical distance between $r_0$ and
$r_+$ is
\be\delta'\approx r_+ \sqrt{\epsilon_0}\approx
\frac{\ell_p}{7.27}\,,\label{ds.rst}\ee
which is also independent of the properties of the black hole.

\item Thirdly, the energy at the cutoff is approximately
\be\cE_{N_0,m}\approx\frac{N_0}{r_+}\sqrt{\frac{\epsilon_0}2}\;
\approx\frac{0.22}{r_+}\,.\label{e.rst}\ee
\end{itemize}

Since  the cutoff energy (\ref{e.rst}) is calculated by using the
canonical time $t$, it is natural to interpret it as the energy
measured by an observer at the spatial infinity. Translating back
to the boundary, this means that the cutoff energy is
\be\cE'_{N_0,m}\approx\frac{\cE_{N_0,m}}{\sqrt{\epsilon_0}}
\sim\frac1{\ell_p}\,,\label{e.rst2}\ee
which is at the Planck scale. This result fits well with our naive
expectations and it also, interestingly, suggests the presence of
a true firewall.

\section{Summary}

Apart from some very special cases (see, e.g. \cite{Sen:2007qy}),
most of our understanding of black hole entropy comes without an
explicit knowledge of the physical entities that actually carry
the thermodynamical properties of the black hole. In view of the
fact that the horizon acts as a (at least, causal) boundary to the
spacetime, we ask what happens if it is in fact a physical and
dynamical boundary to the spacetime, and then we study its
contribution to the black hole thermodynamics.

The dynamics of a physical boundary is naturally governed by the
boundary action, which can be identified by requiring that the
variational principle is well defined. If the bulk theory is
Einstein-Hilbert, then the boundary action is given by the
Gibbons-Hawking-York term. Using the Kerr black hole as an
example, we find that the boundary action on the average location
of the physical horizon effectively describes a massless
Klein-Gordon field. Quantum mechanically, this inevitably leads to
fluctuations of the boundary. We note that only large wave number
modes have significant contributions to the effective action. The
full spectrum of the system can be found in the small rotation
limit of the black hole.

We then look at the contribution of the boundary to the black hole
thermodynamics. It turns out that the angular momentum of the
boundary matches exactly with that of the black hole. It is then
tempting to suggest that all charges of the black hole are carried
by the horizon. We use this as an assumption to fix the two
unknown parameters in our model. As a result, the lattice spacing
of the putative substructures of spacetime along the boundary is
found to be (\ref{a.rst}), the physical distance from the real
boundary to the background horizon $r_+$ is found to be
(\ref{ds.rst}). We also determined the cutoff energy of the
excitations in (\ref{e.rst}).

Phenomenologically, our simple model is already capable of a
detailed explanation of the statistical origin of the black hole
entropy. At the deeper level, however, the model relies on two
unusual assumptions which require further investigation. Let's
finish by listing them:
\begin{itemize}
\item Firstly, our model assumes that the black hole horizon is a
{\it physical} boundary to the spacetime. If there is indeed
substructures to the spacetime, then the existence of a boundary
means that the substructures are in two different phases across
the horizon.

Although it is widely accepted that no one can report to us any
information from behind the horizon of a black hole, people also
often assume that one can pass through the horizon unimpeded (see,
however, \cite{Braunstein:2009my,Almheiri:2012rt}). Our assumption
implies that the spacetime behind the horizon is significantly
different from what is predicted by the usual black hole metric.

\item Secondly, the zero point energy of the quantum modes
(\ref{H.rst3}) plays an indispensable role in the calculation. The
problem of zero point energy is not so crucial for a theory in
flat spacetime. And its role in the case of gravity is not clear.
Our model suggests that black holes may provide the first evidence
that zero point energy is physically relevant in curved spacetime.

It is possible that the horizon is only one of the many physical
entities that contribute to the black hole thermodynamics. If we
do not induce the contribution from the zero point energy, other
sources will have to be included to make up the total black hole
mass. In that case, the simplicity pertained to the present model
will be lost. In particular, it is possible that one may need even
more unusual assumptions in order to explain the unusual relation
between the black hole mass and entropy.
\end{itemize}

\section*{Acknowledgement}

The author thanks Maciej Trzetrzelewski for many discussions
during an earlier related attempt.

\end{document}